\documentclass[twocolumn,showpacs,10pt]{revtex4}

\usepackage{graphicx}
\usepackage{dcolumn}
\usepackage{bm}
\usepackage{color}
\usepackage{amssymb,amsmath}

\input epsf

\begin{document}

\title{\Large Accretion and Evaporation of Modified Hayward Black Hole}

\author{\bf Ujjal Debnath\footnote{ujjaldebnath@gmail.com ,
ujjal@iucaa.ernet.in}}

\affiliation{Department of Mathematics, Indian Institute of
Engineering Science and Technology, Shibpur, Howrah-711 103,
India.\\}

\date{\today}

\begin{abstract}
First we have assumed the most general static spherically
symmetric black hole metric. The accretion of any general kind of
fluid flow around the black hole have been investigated. The
accretion of fluid flow around the modified Hayward black hole
have been analyzed and we then calculated the critical point,
fluid 4 velocity and velocity of sound during accretion process.
Also the nature of the dynamical mass of black hole during
accretion of fluid flow and taking into consideration of Hawking
radiation from black hole i.e., evaporation of black hole have
been analyzed.
\end{abstract}

\pacs{04.70.Bw, 04.70.Dy, 98.80.Cq}
\maketitle

\section{Introduction}

At present we live in a Universe which is expanding and the
expansion rate is increasing i.e, the Universe is accelerating
which was confirmed by recent Supernova type Ia observations
\cite{Riess,Perl}. The large scale structure \cite{Teg,Abaz},
cosmic microwave background radiation \cite{Sper1}, WMAP
observations \cite{Sper2,Briddle,Benn} also support this
acceleration of the Universe. This acceleration is caused by some
unknown matter which produce strong sufficient negative pressure
(with positive energy density), known as {\it dark energy}. The
present Universe occupies $\sim$ 4\% ordinary matter, $\sim$ 74\%
dark energy and $\sim$ 22\% dark matter. Since dark energy and
dark matter are main two components in our universe, such that the
present dark energy and dark matter densities are $7.01\times
10^{-27} kg/m^{3}$ and $2.18\times 10^{-27} kg/m^{3}$,
respectively. Most simplest candidate of dark energy is the
cosmological constant $\Lambda$ which obeys the equation of state
EoS $p=w\rho$ with EoS parameter $w=-1$ \cite{Pad,Sahni}. Another
candidates of dark energy are quintessence (where EoS parameter
satisfies $-1<w<-1/3$) \cite{Peeble,Cald} and phantom (where EoS
parameter satisfies $w<-1$) \cite{Cald1}. Till now there are lot
of dark energy models have been considered. A brief
review of dark energy models is found in the ref. \cite{Cop}.\\

A condensed object (e.g. neutron stars, black holes, etc.)
surrounded by a fluid can capture particles of fluid that pass
within a certain distance from the condensed object. This
phenomena is termed as {\it accretion} of fluid by condensed
objects. In Newtonian theory of gravity, the problem of accretion
of matter onto the compact object was first formulated by Bondi
\cite{Bondi}. Michel \cite{Michel} first obtained analytic
relativistic accretion (of gas) solution onto the static
Schwarzschild black hole. Such accretion processes are candidates
to mechanisms of formation of supermassive black holes (SMBH) in
the center of most active galaxies \cite{Moffat}. In particular,
it should follow some analogies with the process proposed by
Salpeter et al \cite{Merr} where galaxies and quasars could get
some of their energy from processes of accretion. Using this
accretion procedure, Babichev et al \cite{Babichev,Babichev1}
formulated the accretion of phantom dark energy onto a static
Schwarzschild black hole and shown that static Schwarzschild black
hole mass will gradually decrease due to strong negative pressure
of phantom energy and finally all the masses tend to zero near the
big rip singularity. Sun \cite{Sun} discussed phantom energy
accretion onto a Black hole in the cyclic universe. Jamil
\cite{Jamil} has investigated accretion of phantom like modified
variable Chaplygin gas onto Schwarzschild black hole. Phantom
energy accretion by stringly charged black hole has been discussed
by Sharif et al \cite{Sharif1}. Dark matter and dark energy
accretion onto static black hole has been discussed by Kim et al
\cite{Kim}. Also the accretion of dark energy onto the more
general Kerr-Newman black hole was studied by Madrid et al
\cite{Pedro}. The new variable modified Chaplygin gas and
generalized cosmic Chaplygin gas dark energy accretions and onto
Kerr-Newman black hole and their features were studied Bhadra et
al \cite{Bhadra}. Several authors
\cite{Nayak,Dwiv,Lima,Sharif,Mar,Rod,Abhas1,Mar1} have discussed
the accretions of various components of dark energy onto several
types of black holes.\\

In the present work, first we assume the most general static
spherically symmetric black hole metric in section II. The
accretion of any general kind of fluid flow around the black hole
will be investigated. The accretion of fluid flow around the
modified Hayward black hole will be analyzed in section III and we
then calculate the critical point, fluid 4 velocity and velocity
of sound during accretion process. Also the nature of the
dynamical mass of black hole during accretion of fluid flow and
taking into consideration of Hawking radiation from black hole
i.e., evaporation of black hole will be analyzed in section IV.
Finally we shall draw fruitful discussions about accretion of
fluids upon modified Hayward black hole in section V.

\section{Accretion Phenomena of general static spherically symmetric Black Hole}
First we consider general static spherically symmetric metric
given by
\begin{eqnarray}\label{1}
ds^2=-A(r)dt^{2}+\frac{1}{B(r)}~dr^{2}+r^{2}(d\theta^{2}+sin\theta
d\phi^{2})
\end{eqnarray}
where $A(r)>0$ and $B(r)>0$ are functions of $r$ only. We can
choose $A(r)$ and $B(r)$ in such a way that the above metric
represents a black hole metric. Let us assume $M$ is the mass of
the black hole. For instance, if $A(r)=B(r)=1-\frac{2M}{r}$~, the
above metric represents Schwarzschild black hole.

The energy-momentum tensor for the fluid is given by
\begin{equation}\label{2}
T_{\mu\nu}=(\rho+p)u_{\mu}u_{\nu}+pg_{\mu\nu}
\end{equation}
where $\rho$ and $p$ are the energy density and pressure of the
fluid. The four velocity vector of the fluid flow is given by
$u^{\mu}=\frac{dx^{\mu}}{ds}=(u^{0},u^{1},0,0)$ where $u^{0}$ and
$u^{1}$ are the non-zero components of velocity vector satisfying
$u_{\mu}u^{\mu}=-1$. This implies
$g_{00}u^{0}u^{0}+g_{11}u^{1}u^{1}=-1$. So we can obtain
$(u^{0})^{2}=\frac{(u^{1})^{2}+B}{AB}$ and let the radial velocity
of the flow $u^{1}=u$, so we have
$u_{0}=g_{00}u^{0}=\sqrt{\frac{A}{B}}\sqrt{u^{2}+B}$. Here
$\sqrt{-g}=\sqrt{\frac{A}{B}}~r^{2}sin\theta$. From above equation
(\ref{2}), we obtain $T_{0}^{1}=(\rho+p)u_{0}u$. It is assumed
that $u<0$ for inward flow of the fluid towards the black hole.\\

In the fluid flow, we may assume that the fluid is dark matter or
any kind of dark energy. A proper dark-energy accretion model for
static spherically symmetric black hole should be obtained by
generalizing the Michel's theory \cite{Michel}. In the dark energy
accretion onto Schwarzschild black hole, Babichev et al
\cite{Babichev,Babichev1} have performed the above generalization.
We shall follow now the above procedure in the case of static
spherically symmetric black hole. The relativistic Bernoulli's
equation (the time component) of the energy-momentum conservation
law $T^{\mu\nu}_{;\nu}=0$, we obtain
$\frac{d}{dr}~(T_{0}^{~1}\sqrt{-g})=0$ which provides the first
integral, $(\rho+p)u_{0}u^{1}\sqrt{-g}=C_{1}$, that simplifies to
\begin{eqnarray}\label{3}
ur^{2}M^{-2}(\rho+p)\frac{A}{B}\sqrt{u^{2}+B}=C_{1}
\end{eqnarray}
where $C_{1}$ is the integration constant which has the dimension
of the energy density. Moreover, the energy flux equation can be
derived by the projection of the conservation law for
energy-momentum tensor onto the fluid four-velocity, i.e., $u_\mu
T^{\mu \nu}_{;\nu}=0$, which gives
$u^{\mu}\rho_{,\mu}+(\rho+p)u^{\mu}_{;\mu}=0$. From this, we
obtain
\begin{eqnarray}\label{4}
ur^{2}M^{-2}\sqrt{\frac{A}{B}}~exp\left[\int_{\rho_{\infty}}^{\rho_{h}}\frac{d\rho}{\rho+p(\rho)}
\right]=-C
\end{eqnarray}
where $C$ is integration constant (energy flux onto the black
hole) and the associated minus sign is taken for convenience. Also
$\rho_{h}$ and $\rho_{\infty}$ represent the energy densities at
the black hole horizon and at infinity respectively. Combining
equations (\ref{3}) and (\ref{4}), we obtain,
\begin{eqnarray}\label{5}
(\rho+p)\sqrt{u^{2}+B}\sqrt{\frac{A}{B}}~exp\left[-\int_{\rho_{\infty}}^{\rho_{h}}\frac{d\rho}{\rho+p(\rho)}
\right]=C_{2}
\end{eqnarray}
where, $C_{2}=-C_{1}/C=\rho_{\infty}+p(\rho_{\infty})$. The
equation of mass flux $J^{\mu}_{;\mu}=0$ is given by
$\frac{d}{dr}~(J^{1}\sqrt{-g})=0$, which integrates to $\rho
u^{1}\sqrt{-g}=A_{1}$ and yields
\begin{eqnarray}\label{6}
\rho ur^{2}M^{-2}\sqrt{\frac{A}{B}}=C_{3}
\end{eqnarray}
where, $C_{3}$ is the integration constant. From (\ref{3}) and
(\ref{6}), we obtain,
\begin{eqnarray}\label{7}
\frac{\rho+p}{\rho}\sqrt{\frac{A}{B}}~\sqrt{u^{2}+B}=\frac{C_{1}}{C_{3}}=C_{4}=\text{
constant}
\end{eqnarray}
Now let us assume,
\begin{eqnarray}\label{8}
V^{2}=\frac{d\ln(\rho+p)}{d\ln\rho}-1
\end{eqnarray}
So from equations (\ref{6}), (\ref{7}) and (\ref{8}), we obtain
\begin{eqnarray*}
\left[V^{2}-\frac{u^{2}}{u^{2}+B}
\right]\frac{du}{u}+\left[-2V^{2}+\frac{1}{2}\left(\frac{A'}{A}
-\frac{B'}{B}\right)\right.
\end{eqnarray*}
\begin{eqnarray}\label{9}
\left.\times (V^{2}+1)r+\frac{rB'}{2(u^{2}+B)}
\right]\frac{dr}{r}=0
\end{eqnarray}

Now if one or the other of the bracketed terms in (\ref{9})
vanishes, we get a turn-around point and in this case, the
solutions will be the double-valued in either $r$ or $u$. There
are only solutions which pass through a critical point that
correspond to material falling into (or flowing out of) the object
with monotonically increasing velocity along with the particle
trajectory. A point where speed of flow is equal to the speed of
sound, such a point is called a {\it critical point}. It is
assumed that the critical point of accretion is located at
$r=r_{c}$ which is obtained by taking the both bracketed terms
(coefficients of $du$ and $dr$) in Eq. (\ref{9}) to be zero. So at
the critical point, we obtain
\begin{eqnarray}\label{10}
V_{c}^{2}=\frac{u_{c}^{2}}{u_{c}^{2}+B(r_{c})}
\end{eqnarray}
and
\begin{eqnarray}\label{11}
\frac{4V_{c}^{2}}{r_{c}}=\left[\frac{A'(r_{c})}{A(r_{c})}-\frac{B'(r_{c})}{B(r_{c})}\right](V_{c}^{2}+1)+
\frac{B'(r_{c})}{u_{c}^{2}+B(r_{c})}
\end{eqnarray}
Here, subscript $c$ denotes the critical value and $u_{c}$ is the
critical speed of flow at the critical point $r_{c}$. From above
two expressions, we have
\begin{eqnarray}\label{12}
u_{c}^{2}=\frac{B'(r_{c})}{2}\frac{A'(r_{c})}{A(r_{c})}\left[\frac{2}{r_{c}}
-\frac{A'(r_{c})}{A(r_{c})}+\frac{B'(r_{c})}{B(r_{c})}
\right]^{-1}
\end{eqnarray}
and
\begin{eqnarray}\label{12a}
V_{c}^{2}=\left[1+2\frac{A(r_{c})}{A'(r_{c})}\frac{B(r_{})}{B'(r_{c})}
\left(\frac{2}{r_{c}}
-\frac{A'(r_{c})}{A(r_{c})}+\frac{B'(r_{c})}{B(r_{c})}\right)\right]^{-1}
\end{eqnarray}
At the critical point $r_{c}$~, the sound speed can be determined
by
\begin{eqnarray}\label{13}
c_{s}^{2}=\left. \frac{d
p}{d\rho}\right|_{r=r_{c}}=\frac{C_{4}V_{c}(V_{c}^{2}+1)}{u_{c}}~\sqrt{\frac{B(r_{c})}{A(r_{c})}}
-1
\end{eqnarray}
The physically acceptable solutions of the above equations may be
obtained if $u_{c}^{2}>0$ and $V_{c}^{2}>0$ which leads to
\begin{eqnarray}\label{14}
A'(r_{c})B'(r_{c})>0 ~~\text{and}~~\frac{2}{r_{c}}
>\frac{A'(r_{c})}{A(r_{c})}-\frac{B'(r_{c})}{B(r_{c})}
\end{eqnarray}
From the above equation we can obtain the bound of $r_{c}$ if $A$
and $B$ are known for several kinds of static black holes.

\section{Accretion Phenomena of Modified Hayward Black Hole}

The static spherically symmetric space-time is described by the
Hayward metric which is obtained by $A(r)=B(r)$ in equation
(\ref{1}) and is given by \cite{Hay}
\begin{eqnarray}\label{17}
ds^2=-B(r)dt^{2}+\frac{1}{B(r)}~dr^{2}+r^{2}(d\theta^{2}+sin\theta
d\phi^{2})
\end{eqnarray}
Here, $M$ is the mass of Hayward black hole and
$B(r)=1-\frac{2Mr^{2}}{r^{3}+2Ml^{2}}$~, where $l$ is a parameter
with dimensions of length (Hubble length) with small scale related
to the inverse cosmological constant $\Lambda$ ($l$ is a
convenient encoding of the central energy density $\frac{3}{8\pi
l^{2}}\sim \Lambda$, assumed positive). Such behavior has been
proposed by Sakharov \cite{Sar,Gli} as the equation of state of
matter at high density and by Markov \cite{Mark,Fro} based on an
upper limit on density or curvature, to be ultimately justified by
a quantum theory of gravity. In the limit $r\rightarrow\infty$,
$B(r)\approx 1-\frac{2M}{r}$ which represents Schwarzschild black
hole, but it becomes de-Sitter black hole as $B(r)\approx
1-\frac{r^{2}}{l^{2}}$ near the center ($r\approx 0$), so it is a
regular space-time without singularity. Thus Hayward black hole is
the simplest regular black hole. Some physical consequences of
Hayward black hole have been discussed by several authors
\cite{Ab0,Lin,Hali}. After that the Hayward metric was modifies
\cite{Lor} by choosing $A(r)=f(r)B(r)$, satisfying the conditions:
it (i) preserves the Schwarzschild behaviour at large $r$, (ii)
includes the 1-loop quantum corrections and (iii) allows for a
finite time dilation between the center and infinity. So the
modified Hayward black hole metric is given by \cite{Lor},
\begin{eqnarray}\label{18}
ds^2=-f(r)B(r)dt^{2}+\frac{1}{B(r)}~dr^{2}+r^{2}(d\theta^{2}+sin\theta
d\phi^{2})
\end{eqnarray}
where
\begin{eqnarray}\label{18a}
B(r)=1-\frac{2Mr^{2}}{r^{3}+2Ml^{2}}~,~f(r)=1-\frac{\alpha\beta
M}{\alpha r^{3}+\beta M}
\end{eqnarray}
with $\alpha,~\beta$ are positive constants. Now from the relation
$A(r)=f(r)B(r)$, we may obtain
\begin{eqnarray}\label{19}
\frac{A'(r)}{A(r)}=\frac{f'(r)}{f(r)}+\frac{B'(r)}{B(r)}
\end{eqnarray}
Also from the expressions of $B(r)$ and $f(r)$ (equation
(\ref{18a})), we get
\begin{eqnarray}\label{20}
\frac{B'(r)}{B(r)}=\frac{2Mr(r^{3}-4Ml^{2})}{(r^{3}+2Ml^{2})[r^{3}+2M(l^{2}-r^{2})]}~,
\end{eqnarray}
\begin{eqnarray}\label{21}
\frac{f'(r)}{f(r)}=\frac{3\alpha^{2}\beta Mr^{2}}{(\alpha
r^{3}+\beta M)[\alpha r^{3}+(1-\alpha)\beta M]}
\end{eqnarray}
Since for outside the horizon, (i) $B(r)>0$ which implies
$r^{3}>2M(r^{2}-l^{2})$ and (ii) $f(r)>0$, so we get,
$r>\left[\frac{\beta(\alpha-1)M}{\alpha} \right]^{\frac{1}{3}}$
with $\alpha>1$. So from equation (\ref{21}), we have $f'(r)>0$.\\

If we assume that the fluid flow accretes upon modified Hayward
black hole, we can calculate the expressions of
$u_{c}^{2},~V_{c}^{2}$ and $c_{s}^{2}$ at the critical point
$r_{c}$. The expressions are given below (using equations
(\ref{12}), (\ref{12a}) and (\ref{13})):
\begin{eqnarray}\label{22}
u_{c}^{2}=\frac{B'(r_{c})}{2}\left(\frac{B'(r_{c})}{B(r_{c})}+\frac{f'(r_{c})}{f(r_{c})}\right)\left(
\frac{2}{r_{c}}-\frac{f'(r_{c})}{f(r_{c})}\right)^{-1}~,
\end{eqnarray}
\begin{eqnarray}\label{23}
V_{c}^{2}=\left[1+2\frac{B(r_{c})}{B'(r_{c})}
\left(\frac{B'(r_{c})}{B(r_{c})}+\frac{f'(r_{c})}{f(r_{c})}\right)^{-1}\left(
\frac{2}{r_{c}}-\frac{f'(r_{c})}{f(r_{c})}\right)\right]^{-1}~,
\end{eqnarray}
\begin{eqnarray}\label{23}
c_{s}^{2}=\frac{C_{4}V_{c}(V_{c}^{2}+1)}{u_{c}\sqrt{f(r_{c})}}-1
\end{eqnarray}
where $B(r)$, $f(r)$ and their derivatives are given in
(\ref{18a}), (\ref{20}) and (\ref{21}) at the point $r=r_{c}$. The
physically acceptable solutions of the above equations may be
obtained if $u_{c}^{2}>0$ and $V_{c}^{2}>0$ which leads to
\begin{eqnarray}\label{24}
B'(r_{c})>0~\text{and}~0<\frac{f'(r_{c})}{f(r_{c})}<\frac{2}{r_{c}}
\end{eqnarray}
From above restrictions, we may get the bounds of $r_{c}$ and that
is ($\alpha>1$):
\begin{eqnarray}\label{25}
r_{c}^{3}>Max\left\{4Ml^{2},\frac{\beta(-4+5\alpha+\sqrt{\alpha(25\alpha-24)})
M}{4\alpha} \right\}
\end{eqnarray}
For example, we assume a fluid flow obeys linear equation of state
$p=w\rho$ ($w=$ constant) accretes upon modified Hayward black
hole. Then we obtain $c_{s}^{2}=w$ and $V_{c}^{2}=0$ and from
(\ref{10}), we obtain $u_{c}=0$. From equations (\ref{22}) and
(\ref{23}), we see that the critical point occurs at the point
$r_{c}=(4Ml^{2})^{\frac{1}{3}}$. For general equation of state
where $w=w(t)$, we obtain $c_{s}^{2}\ne$ constant, $V_{c}^{2}\ne
0$ and $u_{c}^{2}\ne 0$. In this case, it is very difficult to
obtain the critical point $r_{c}$.\\

\section{Changes of Black Hole Mass During Accretion and Evaporation}

The rate of change of mass $\dot{M}$ of the black hole is computed
by integrating the flux of the fluid over the 2-dimensional
surface of the black hole and given by $\dot{M}=-\int T_{0}^{1}dS$
where $dS=\sqrt{-g}d\theta d\phi$. Using equation (\ref{3}), we
obtain the rate of change of mass of black hole as in the
following form:
\begin{eqnarray}\label{25a}
\dot{M}=4\pi CM^{2}(\rho_{\infty}+p(\rho_{\infty}))
\end{eqnarray}
The above result is also valid for any equation of state
$p=p(\rho)$. So the rate of change of mass for the accreting fluid
around the black hole will be
\begin{eqnarray}\label{25b}
\dot{M}_{acc}=4\pi CM^{2}(\rho+p)
\end{eqnarray}
We see that the rate of change of mass for the general spherically
symmetric static black hole due to accretion of fluid flow becomes
exactly similar rate in the case of a Schwarzschild black hole.
From the expression (\ref{25b}) it is to be noted that {\it the
rate of change of mass of any static spherically symmetric black
hole is completely independent of $A(r)$ and $B(r)$}. When some
fluid accretes outside the black hole, the mass function $M$ of
the black hole is considered as a dynamical mass function and
hence it should be a function of time also. So $\dot{M}$ is time
dependent and the increasing or decreasing of the black hole mass
$M$ sensitively depends on the nature of the fluid which accretes
upon the black hole. If $\rho+p<0$ i.e., for phantom dark energy
accretion, the mass of the black hole decreases but if $\rho+p>0$
i.e., for quintessence dark energy accretion, the mass of the black hole increases.\\

We may also assume that the black hole evaporates by Hawking
radiation process. The rate of change of mass for the evaporation
is given by
\begin{eqnarray}\label{27}
\dot{M}_{eva}=-\frac{D}{M^{2}}
\end{eqnarray}
where $D>0$ is a constant whose value depends on the model
\cite{Cline}. Now due to accretion of fluid flow and evaporation
of mass of black hole, we get the rate of change of mass of black
hole as
\begin{eqnarray}\label{28}
\dot{M}=\dot{M}_{acc}+\dot{M}_{eva}=4\pi
CM^{2}(\rho+p)-\frac{D}{M^{2}}
\end{eqnarray}
For accretion scenario, the changes of mass of black hole
completely depends on the nature of the fluid accretes. But for
evaporation process, the change of mass of black hole is
independent of the nature of fluid, because this is internal
process. In fact, when the accretion fluid is only the
cosmological constant ($p=-\rho$), the mass of black hole for only
accretion scenario is always same throughout time evolution. Only
in accretion process, the mass of black hole increases for normal
fluid and quintessence type dark energy fluid and decreases for
phantom dark energy. But due to accretion as well as evaporation,
$\dot{M}>0$ for $M^{4}>\frac{D}{4\pi C(\rho+p)}$ and $\dot{M}<0$
for $M^{4}<\frac{D}{4\pi C(\rho+p)}$ for normal fluid and
quintessence type dark energy, but for phantom energy, black hole
mass always decreases ($\dot{M}<0$). Thus evaporation supports the
decreasing of the mass of black hole with some restrictions of
minimum values of mass of black hole.\\

\section{Discussions and Concluding Remarks}

First we have assumed the most general static spherically
symmetric black hole metric. The accretion of any general kind of
fluid flow around the black hole have been investigated. For this
general kind of static black hole, the critical point, velocity of
sound, fluid 4 velocity have been calculated and shown that these
value depend completely on the metric coefficients. Next, the
accretion of fluid flow around the modified Hayward black hole
have been analyzed and we then calculated the critical point,
fluid 4 velocity and velocity of sound during accretion process.
We can mention that for outside the horizon, (i) $B(r)>0$ which
implies $r^{3}>2M(r^{2}-l^{2})$ and (ii) $f(r)>0$, so we get,
$r>\left[\frac{\beta(\alpha-1)M}{\alpha} \right]^{\frac{1}{3}}$
with $\alpha>1$ and also $f'(r)>0$. For physical region of
accretion have been found and the bounds of critical point have
been generated and the bound of critical point is
$r_{c}^{3}>Max\left\{4Ml^{2},\frac{\beta(-4+5\alpha+\sqrt{\alpha(25\alpha-24)})
M}{4\alpha} \right\}$. When the perfect fluid satisfying linear
equation of state $p=w\rho$ ($w=$ constant) accretes upon modified
Hayward black hole we have obtained $c_{s}^{2}=w$, $u_{c}=0$ and
$V_{c}^{2}=0$. In this accretion process, we have seen that the
critical point occurs at the point
$r_{c}=(4Ml^{2})^{\frac{1}{3}}$. Also the nature of the dynamical
mass of black hole during accretion of fluid flow and taking into
consideration of Hawking radiation from black hole i.e.,
evaporation of black hole have been analyzed. Only in accretion
process, the mass of black hole increases for normal fluid and
quintessence type dark energy fluid and decreases for phantom dark
energy. But due to accretion as well as evaporation, $\dot{M}>0$
for $M^{4}>\frac{D}{4\pi C(\rho+p)}$ and $\dot{M}<0$ for
$M^{4}<\frac{D}{4\pi C(\rho+p)}$ for normal fluid and quintessence
type dark energy, but for phantom energy, black hole mass always
decreases ($\dot{M}<0$). Thus evaporation supports the decreasing
of the mass of black hole with some restrictions of
minimum values of mass of black hole.\\

{\bf Acknowledgement:}\\

The author is thankful to IUCAA, Pune,
India for warm hospitality where the work was carried out.\\

\end{document}